# Layer-by-layer epitaxial growth of polar FeO(111) thin films on MgO(111)

Jacek Gurgul[1,*], Ewa Młyńczak[1], Nika Spiridis[1], Józef Korecki[1,2]

[1]*Jerzy Haber Institute of Catalysis and Surface Chemistry, Polish Academy of Sciences, Niezapominajek 8, 30-239 Kraków, Poland*

[2]*Faculty of Physics and Applied Computer Science, AGH University of Science and Technology, al. Mickiewicza 30, 30-059 Kraków, Poland*

*Corresponding author (ncgurgul@cyf-kr.edu.pl)
Jerzy Haber Institute of Catalysis and Surface Chemistry, Polish Academy of Sciences, Niezapominajek 8, 30-239 Kraków, Poland
Tel. +48 126395116; Fax. +48 124251923

e-mail addresses:    UUncgurgul@cyf-kr.edu.pl
ncmlyncz@cyf-kr.edu.pl
ncspirid@cyf-kr.edu.pl
korecki@agh.edu.pl

**Abstract**

We report on the structural properties of epitaxial FeO layers grown by molecular beam epitaxy on MgO(111). The successful stabilization of polar FeO films as thick as 16 monolayers (ML), obtained by deposition and subsequent oxidation of single Fe layers, is presented. FeO/MgO(111) thin films were chemically and structurally characterized using low-energy electron diffraction, Auger electron spectroscopy and conversion electron Mössbauer spectroscopy (CEMS). Detailed *in situ* CEMS measurements as a function of the film thickness demonstrated a size-effect-induced evolution of the hyperfine parameters, with the thickest film exhibiting the bulk-wüstite hyperfine pattern. *Ex situ* CEMS investigation confirmed the magnetic ordering of the wüstite thin film phase at liquid nitrogen temperature.

# 1. Introduction

The latest advances in the description of polar oxide surfaces have shed light on the physical properties of the surfaces and interfaces of ionic materials [1]. However, our understanding of low-dimensionality effects in such materials is still very poor. Theoretical predictions and experimental results have shown that bulk-terminated polar surfaces are not stable because alternating layers of oppositely charged ions produce a large dipole moment perpendicular to the surface, which results in a diverging electrostatic potential. Two mechanisms provide solutions to this infinite energy problem: the surface could facet into neutral planes or it could reconstruct to balance its surface charge [1, 2]. In contrast to the surfaces of single crystals, non-reconstructed polar surfaces can be produced in thin films below a critical thickness [1 and references therein]. From this perspective, the preparation and study of stable non-reconstructed polar surfaces are of great interest. On the other hand, the stabilization of orientations, which do not occur naturally when cleaving a bulk material, opens the way to the fabrication of artificial structures for controlled catalysis or nanostructures for electronic and magnetic applications.

The simple cubic rock-salt structure is one of the most stable structures for highly ionic solids, such as metal monoxides (e.g., MgO, CoO, FeO, etc.). It consists of two interpenetrating fcc lattices of anions and cations. While the {001} faces are neutral and can be easily stabilized by epitaxial growth [3], obtaining the polar plane orientation of the lowest index {111} is challenging. A crystal cut along (111) presents alternating layers of metal and oxygen ions. Because of charge compensation, the surface tends to reconstruct or facet. The most common stable surface configuration is found to be the octopolar (2×2) reconstruction [4 and references therein]. This reconstruction is obtained by removing 75 % of the atoms in the outermost layer and 25 % in the layer beneath, which produces {100} facets. Polar surface stabilization mechanisms may be altered in ultrathin polar films and nanostructures because of their reduced dimensionality [5]. Additionally, in these systems, surface relaxation, diffusion of atoms, filling of surface states or covalency modifications are possible stabilizing processes. However, these mechanisms have not yet been quantitatively assessed, and there is a need for experimental work in these fields.

We have chosen to study FeO because of its importance in basic research, as well as in technological applications. Iron monoxide (FeO) adopts the rock salt structure above its Néel temperature ($T_N \approx 198$ K). However, it is well known that FeO is nonstoichiometric, accommodating a cation deficiency by the formation of octahedral iron vacancies and a small number of tetrahedral iron(III) interstitials. These defects tend to aggregate and form tetrahedral units that have been identified by neutron diffraction and Mössbauer spectroscopy [6, 7]. The bulk magnetic properties of wüstite $Fe_yO$ are complex; it is an antiferromagnet with an exact Néel temperature that depends on the value of y [6, 8]. Below the magnetic ordering temperature, it undergoes a rhombohedral distortion, and its iron spins align along the [111] direction of the unit cell, forming antiferromagnetically coupled alternate (111) iron ferromagnetic sheets [9].

Although iron oxide thin films prepared by different methods, ranging from the direct oxidation of metallic Fe(110) and (001) surfaces to the deposition and oxidation of Fe layers on many substrates, are widely investigated (for a review, see [10]), the literature on polar FeO(111) films is scarce. Most studies consider ultrathin FeO(111) films prepared on Pt(111) [11]. There are several experiments showing that FeO(111) can be obtained by oxidation of the Fe(110) surface [12 and references therein], and to our knowledge, there has been only one example of the growth of FeO(111) monolayers on oxide surfaces [13]. In particular, there are no published papers related to FeO/MgO(111) systems.

In this paper, we report the successful preparation of polar FeO thin films on MgO(111). Because of the layer-by-layer technique, we were able to stabilize an epitaxial FeO film as thick as 16 ML. Every step of the preparation sequences was accompanied by detailed structural studies with low electron energy diffraction (LEED), Auger electron spectroscopy (AES) and conversion electron Mössbauer spectroscopy (CEMS). Compared to standard surface sensitive characterization methods, CEMS has the advantage of probing deeper layers (down to 100 nm) with a monolayer resolution, as well as their local structure and symmetry [3, 14].

## 2. Material and methods

The experiments were performed in a multi-chamber ultrahigh vacuum (UHV) system (base pressure $2\times10^{-10}$ mbar) equipped with facilities for the growth of epitaxial films, along with their structural and chemical characterizations and *in situ* CEMS measurements. The preparation chamber contains a molecular beam epitaxy (MBE) system, including an evaporator for iron that is 95 % enriched with the $^{57}Fe$ isotope, a quartz monitor to control the deposition rate, a LEED/AES spectrometer for fast sample characterization and a MgO evaporator. A MgO(111) polished single crystal ($10\times10\times1$ mm$^3$) was used as the substrate. The substrate was annealed at 903 K for 3 h at UHV, and a homoepitaxial 30 Å layer of MgO was deposited at 723 K in three steps (10 Å each) and annealed at 823 K (30 minutes) in oxygen atmosphere ($5 \times 10^{-9}$ mbar) to improve the quality of the surface. Single monolayers of FeO were grown by the deposition of an $^{57}$Fe monolayer at normal incidence at RT and subsequent oxidation in $O_2$ ($5 \times 10^{-8}$ mbar, 10 L) at 543 K, followed by UHV annealing at 873 K for 30 minutes. The FeO monolayer deposition procedure was repeated 16 times, which resulted in the 16 ML FeO film. After every monolayer preparation cycle, the surface structure and composition of the thin films were monitored by LEED and AES.

The *in situ* CEMS measurements were performed at selected preparation stages at room temperature using an UHV spectrometer, similar to that described previously [15] and a standard 100 mCi Mössbauer $^{57}$Co(Rh) γ-ray source. The *in situ* spectra were taken at a fixed angle of 36° between the direction of the γ-ray propagation and the sample normal. For low temperature (LT) *ex situ* CEMS measurements, the sample was capped with a 50 Å protective MgO film. The LT CEMS experiments were performed at 80 K in a separate UHV system [16] equipped with a liquid nitrogen stationary cryostat.

The Mössbauer spectra were analyzed numerically by fitting a hyperfine parameter distribution (HPD) using the Voigt-line-based method of Rancourt and Ping [17]. In this method, the HPD for a given crystal site corresponding to similar structural, chemical and magnetic properties is constructed by a sum of Gaussian components for the quadrupole splitting (QS) distributions and, if necessary, the magnetic hyperfine field $B_{hf}$ distributions. The isomer shift (IS) can be linearly coupled to the primary hyperfine parameters (QS, $B_{hf}$). The isomer shift values are quoted relative to α-Fe at room temperature.

## 3. Results and discussion

The Auger spectrum induced by 1.7 keV electrons showed small amounts of carbon (KE = 270 eV) and calcium (KE = 291 eV) at the polished MgO(111) crystal surface (Fig. 1). To obtain a cleaner surface, homoepitaxial MgO buffer layers were deposited. The deposition of 30 Å of MgO clearly resulted in a calcium-free surface. However, small traces of carbon were still present at the surface. The oxygen KLL Auger electron peak at KE = 508 eV with respect to the Fermi level is characteristic for the uncharged surface of MgO [18].

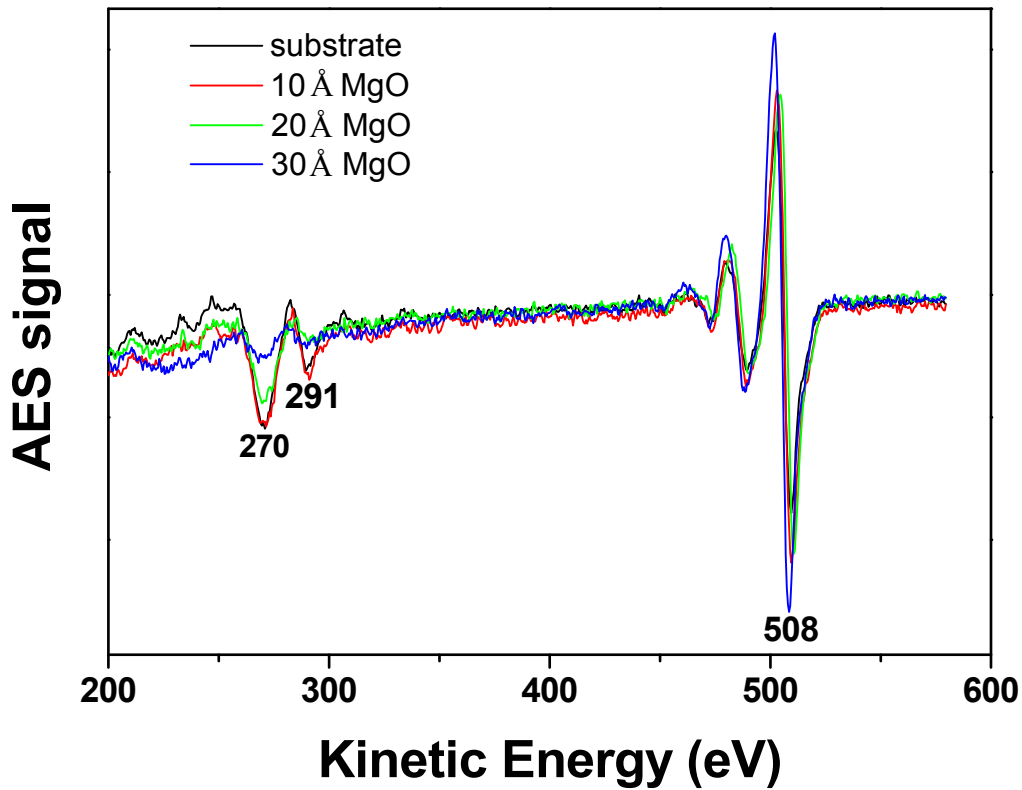

Fig. 1. The AES spectrum for annealed MgO(111) substrate before and after covering with homoepitaxial layers of varying thickness.

Because of the insulating character of MgO, LEED patterns could be obtained only for high electron energies. Figure 2 shows the LEED patterns recorded for a primary electron energy of 190 eV before and after the deposition of 30 Å of MgO. Both patterns exhibit a (1×1) hexagonal symmetry, confirming the homoepitaxial growth of MgO. The spots are less sharp in the case of the surface with an additional 30 Å of MgO, most likely because of stronger charging or slightly poorer structural long-range order.

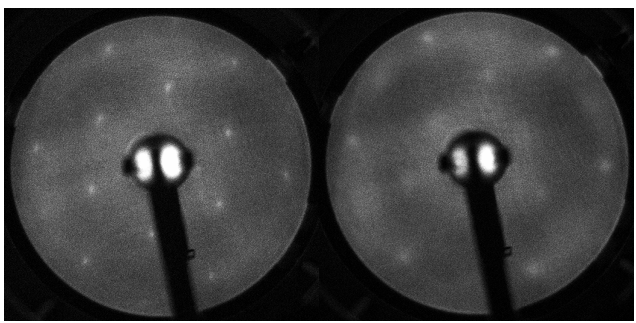

Fig. 2. The LEED patterns at the primary electron energy 190 eV for the MgO(111) substrate after annealing (left) and for 30 Å of homoepitaxial MgO layer (right).

The deposition of the thinnest FeO layers drastically changed the electrical state of the samples, making LEED observation possible down to the lowest energies, which demonstrates the much lower energy gap of FeO compared to MgO.

Figure 3 compares the LEED patterns obtained from the FeO(111) films of different thicknesses. All patterns exhibit a (1x1) hexagonal symmetry that confirms epitaxial growth through the entire FeO film. The orientation of the FeO films is expected to be $(111)_{FeO}//(111)_{MgO}$ because of the small mismatch of the lattice parameters, and we can conclude that the FeO follows the orientation of the MgO substrate with high fidelity.

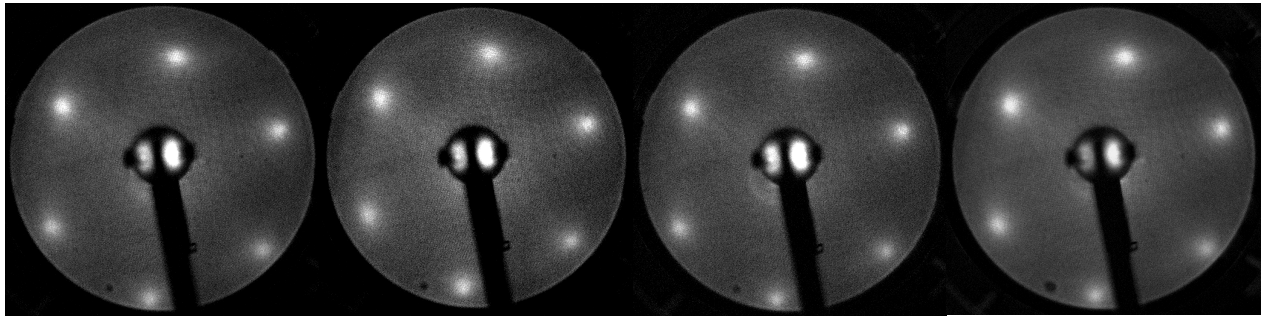

Fig. 3. LEED patterns at the primary electron energy 67 eV of 2, 4, 8 and 16 ML thick FeO, from left to right, respectively.

Remarkably, the metallic Fe monolayers produced very poor diffraction patterns with visible 3-fold symmetry (not shown here), and only the oxidation process caused meaningful improvement of the long range order of the surface. The brightness and sharpness of the spots are comparable for all cases, showing the good crystalline quality of the films and their uniform growth mode; however, the broadening of the diffraction spots is an indication of a grainy film structure. It is well known that the initial growth of iron on MgO is island-like [19], and this adsorbate morphology is transmitted to the oxide film. Under the above-mentioned conditions, there is no appearance of {100} facets, which can be easily understood in view of the low dimensionality of the oxide structure [5].

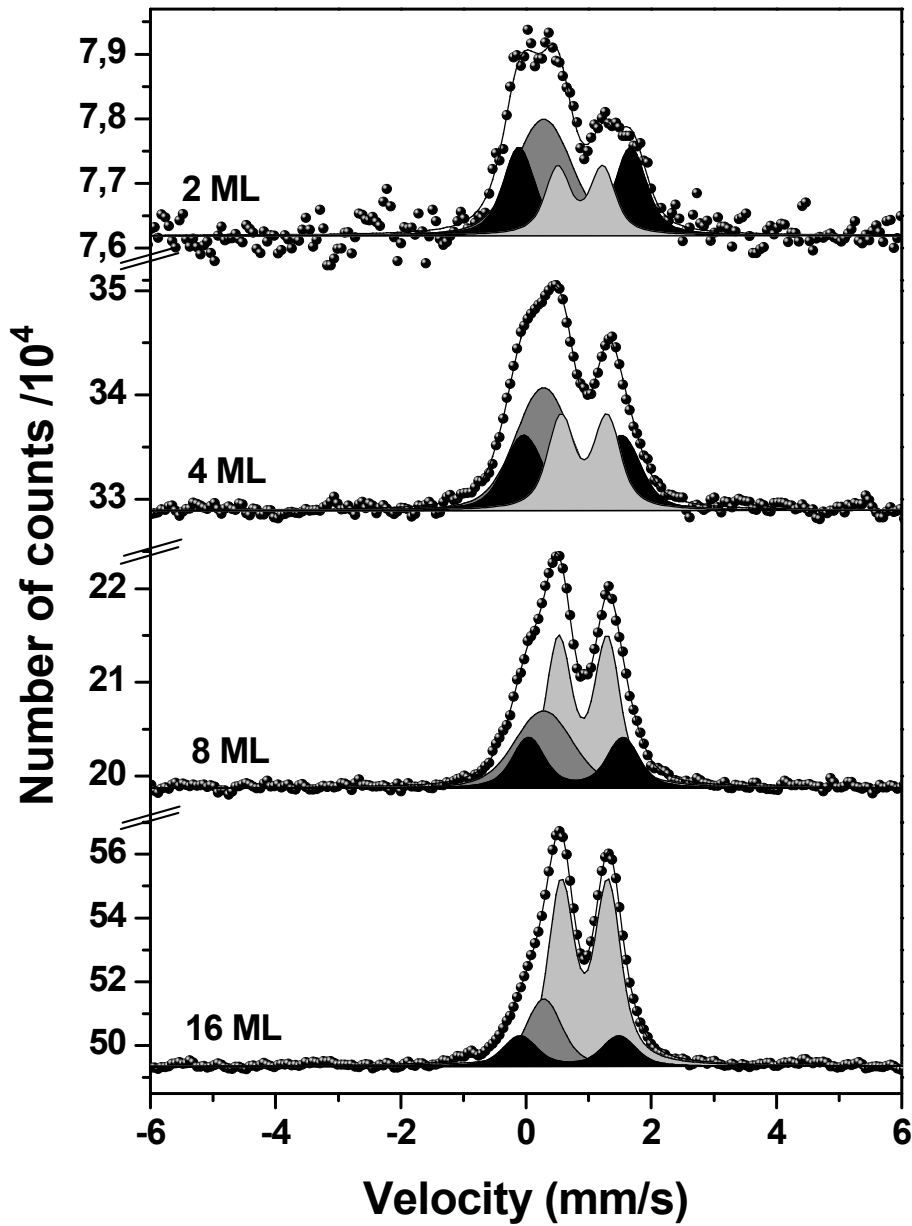

Fig. 4. *In situ* room temperature $^{57}$Fe-CEMS spectra for FeO films on MgO(111) as a function of the FeO thickness.

The room temperature *in situ* CEMS spectra of the FeO(111) films at different growth stage layers are presented in Fig. 4. The apparent asymmetric doublets are qualitatively similar to the well-known bulk-Fe$_y$O spectra in the paramagnetic state [6, 7], but detailed numerical analysis reveals quantitative differences that are not surprising when the possible deviation from stoichiometry and low dimensionality of our samples are considered. Under ambient conditions, FeO crystallizes in the cubic rock-salt structure. The Mössbauer spectrum should nominally contain one singlet corresponding to $Fe^{2+}$ in the octahedral site, but the room temperature spectra show several singlets and doublets because of the presence of undistorted octahedral $Fe^{2+}$ sites, octahedral $Fe^{2+}$ sites associated with vacancies and complex defect clustering and also $Fe^{3+}$ in octahedral and tetrahedral positions [20]. To accurately and consistently describe our spectra, at least two doublets and one single line were required, with different area ratios depending on the film thickness. The hyperfine parameters of the fitted components are listed in Table 1. Both doublets are characterized by a large isomer shift, which is typical for $Fe^{2+}$. The doublet with the larger IS (~ 1.0 mm/s) and smaller QS (~ 0.75 mm/s), which is shaded light gray in Fig. 4, is attributed to the undistorted octahedral $Fe^{2+}$ sites. This assumption is consistent with the values of the hyperfine parameters, which are close to those of the bulk [6]; it is also consistent with the increasing peak intensity with increasing film thickness. From this trend, we make the obvious assumption that the inner layers of the films present more bulk-like behavior than the surface and interface layers. Consequently, the lower IS doublet, which is shaded black in Fig. 4 and decreases in intensity with increasing film thickness, can be attributed to the defect-associated octahedral $Fe^{2+}$ sites. The large QS value of this spectral component (~ 1.50 mm/s) can be explained by low local symmetry and possible thin film distortion, which also contribute to the electric field gradients in this nominally cubic compound. The larger distribution of QS for this component also supports the above interpretation, although it is also plausible that interstitial tetrahedral $Fe^{3+}$ sites, considered to be a separate component in some models of bulk $Fe_yO$ [6], contribute to the broad distribution of QS.

The single line with a small isomer shift, shaded dark gray in Fig. 4, can be attributed to octahedral $Fe^{3+}$ ions, which are related to cation vacancies in the FeO film. The observed value of IS = 0.39 mm/s is typical of the $3d^5$ state [6]. In our fits, we used a broad single line for the iron(III) line rather than a quadrupole doublet. This substitution is reasonable because the high-spin iron(III) has a $^6A_{1g}$ ground state and, hence, only a very small valence contribution to the electric field gradient. The broadening of the line is probably a result of the significant lattice contribution to the EFG, but the low dimensionality of the film can also play an important role.

Table 1. Hyperfine parameters derived from the room-temperature CEMS spectra for FeO films on MgO(111) as a function of the film thickness. Isomer shift (IS) is given in mm/s relative to α-Fe foil. $<QS> = 1/2e^2qQ$ (in mm/s) and σ (in mm/s) are the average quadrupole splitting and the Gaussian width of the QS distribution of the given spectral component, respectively.

|  | Iron(II) | | | | | | | | Iron(III) | | | | |
|---|---|---|---|---|---|---|---|---|---|---|---|---|---|
|  | Central doublet | | | | Outer doublet | | | | Singlet | | | | |
|  | IS | <QS> | σ | % A[a] | IS | <QS> | σ | % A | IS | <QS> | σ | % A | $\chi^2$ |
| 2 ML | 0.97 | 0.72 | 0.22 | 23.2 | 0.89 | 1.79 | 0.34 | 36.6 | 0.39* | 0* | 0.76 | 40.2 | 0.50 |
| 4 ML | 1.03 | 0.74 | 0.25 | 30.1 | 0.85 | 1.56 | 0.45 | 32.5 | 0.39* | 0* | 0.75 | 37.4 | 0.56 |
| 8 ML | 1.02 | 0.78 | 0.24 | 51.8 | 0.90 | 1.51 | 0.35 | 21.3 | 0.39* | 0* | 0.77 | 26.9 | 0.54 |
| 16 ML | 1.04 | 0.75 | 0.24 | 68.8 | 0.80 | 1.58 | 0.38 | 14.4 | 0.39* | 0* | 0.41 | 16.8 | 0.90 |

[a] The percentage area of spectra components.

* Parameters constrained during the fitting procedure.

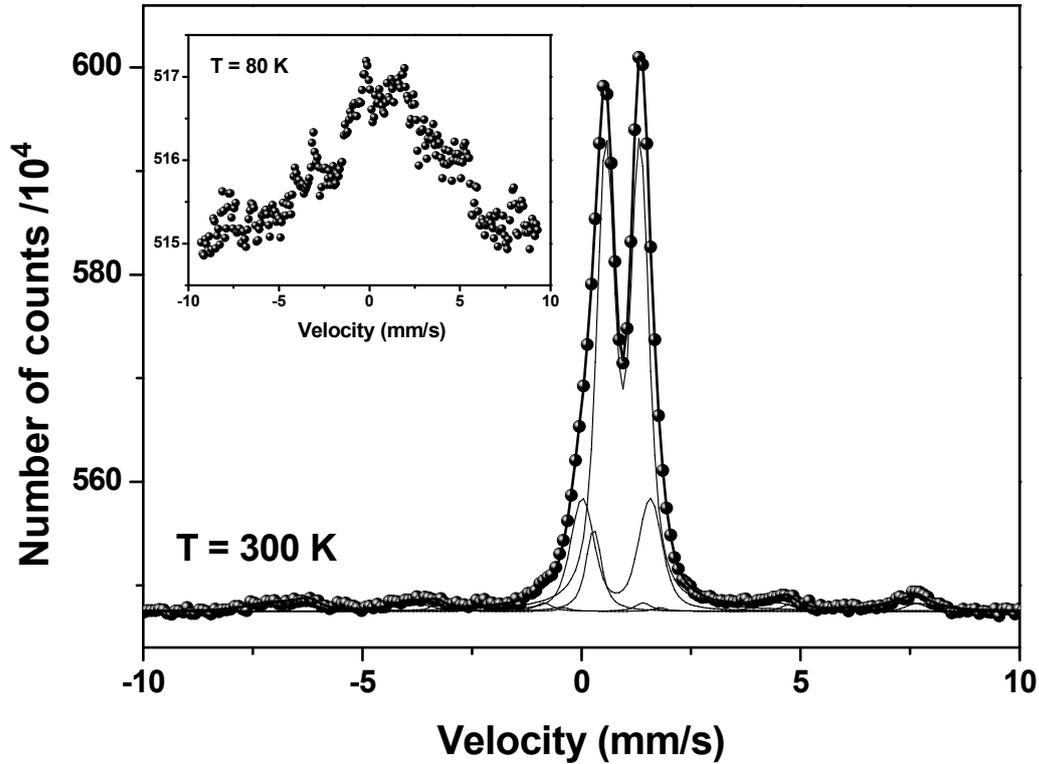

Fig. 5. *Ex situ* CEMS spectra for 16 ML of FeO(111) on MgO(111) at 300 K and 80 K (inset).

The *ex situ* Mössbauer spectra obtained for 16 ML of FeO(111) at 300 and 80 K are shown in Fig. 5. The spectrum at 300 K is very similar to the corresponding spectrum measured *in situ*, but the improved signal-to-noise ratio allows the detection of a small amount (approximately 9 %) of magnetite $Fe_3O_4$ in our sample. The spectrum measured at low temperature is complex, and the spectrum quality excludes an unambiguous interpretation; however, magnetic ordering similar to that observed in the bulk [6] is evident.

## 4. Conclusions

It was shown that epitaxial growth of FeO(111) thin layers on MgO(111) is possible by deposition and subsequent oxidation of single Fe layers. In this case, special attention must be paid to the preparation of the MgO(111) substrate, and a special treatment procedure is necessary during the iron deposition and oxidation. Analysis of the CEMS spectra showed that a FeO wüstite phase is formed. The low-temperature CEMS spectrum measured at 80 K confirmed the presence of a magnetic state of FeO.

**Acknowledgements**

This work was supported in part by the Polish Ministry of Science and Higher Education and by the MPD and Team Programs of the Foundation for Polish Science co-financed by the EU European Regional Development Fund.

**Reference**

[1] C. Noguera, J. Phys.: Condens. Matter 12 (2000) R367.

[2] G.W. Watson, E.T. Kelsey, N.H. de Leeuw, D.J. Harris, S.C. Parker, J. Chem. Soc., Faraday Trans. 92 (1996) 433.

[3] J. Gurgul, K. Freindl, A. Kozioł-Rachwał, K. Matlak, N. Spiridis, T. Ślęzak, D. Wilgocka-Ślęzak, J. Korecki, Surf. Interf. Anal. 42 (2010) 696.

[4] A. Pojani, F. Finocchi, J. Goniakowski, C. Noguera, Surf. Sci. 387 (1997) 354.

[5] J. Goniakowski, C. Noguera, Phys. Rev. B 83 (2011) 115413.

[6] C. Wilkinson, A.K. Cheetham, G.J. Long, P.D. Battle, D.A.O. Hope, Inorg. Chem. 23 (1984) 3136.

[7] N. N. Greenwood, A. T. Howe, J. Chem. Soc., Dalton Trans. 1972, 110.

[8] D.A.O. Hope, A.K. Cheetham, G.J. Long, Inorg. Chem. 21 (1982) 2804.

[9] G.J. Long, D.A.O. Hope, A.K. Cheetham, Inorg. Chem. 23 (1982) 3141.

[10] H.-J. Freund, H. Kuhlenbeck, V. Staemmler, Rep. Prog. Phys. 59 (1996) 283.

[11] W. Weiss, M. Ritter, Phys. Rev. B 59 (1999) 5201

[12] K. Koike, T. Furukawa, Phys. Rev. Lett. 77 (1996) 3921.

[13] S. Gota, E. Guiot, M. Henriot, M. Gautier-Soyer, Phys. Rev. B 60 (1999) 14387.

[14] M. Zając, K. Freindl, T. Ślęzak, M. Ślęzak, N. Spiridis, D. Wilgocka-Ślęzak, J. Korecki, Thin Solid Films 519 (2011) 5588.

[15] N. Spiridis, R.P. Socha, B. Handke, J. Haber, M. Szczepanik, J. Korecki, Catal. Today 169 (2011) 24.

[16]J. Korecki, M. Kubik, N. Spiridis, T. Ślęzak, Acta Phys. Pol. A 97 (2000) 129.

[17] D.G. Rancourt, J.Y. Ping, Nucl. Instrum. Meth. Phys. Rev. B 58 (1991) 85.

[18] M.P. Seah, S.J. Spencer, J. Electron Spec. Relat. Phenom. 109 (2000) 291.

[19] M. Zając, K. Freindl, K. Matlak, M. Ślęzak, T. Ślęzak, N. Spiridis, J. Korecki, Surf. Sci. 601 (2007) 4305.

[20] I.Yu. Kantor, C.A. McCammon, L.S. Dubrovinsky, J. Alloys and Compounds 376 (2004) 5.

**Figure and Table captions**

Fig. 1. The AES spectrum for annealed MgO(111) substrate before and after covering with homoepitaxial layers of varying thickness.

Fig. 2. The LEED patterns at the primary electron energy 190 eV for the MgO(111) substrate after annealing (left) and for 30 Å of homoepitaxial MgO layer (right).

Fig. 3. LEED patterns at the primary electron energy 67 eV of 2, 4, 8 and 16 ML thick FeO, from left to right, respectively.

Fig. 4. *In situ* room temperature $^{57}$Fe-CEMS spectra for FeO films on MgO(111) as a function of the FeO thickness.

Fig. 5. *Ex situ* CEMS spectra for 16 ML of FeO(111) on MgO(111) at 300 K and 80 K (inset).

Table 1. Hyperfine parameters derived from the room-temperature CEMS spectra for FeO films on MgO(111) as a function of the film thickness. Isomer shift (IS) is given in mm/s relative to α-Fe foil. <QS> = $1/2e^2qQ$ (in mm/s) and σ (in mm/s) are the average quadrupole splitting and the Gaussian width of the QS distribution of the given spectral component, respectively.